\documentclass[journal]{IEEEtran}

\usepackage{cite}
\usepackage{amsmath,amssymb,amsfonts,amsthm}
\usepackage{algorithm,algpseudocode}
\usepackage{graphicx,subfigure}
\usepackage{enumitem}
\usepackage{tabularx,multirow,booktabs,makecell}

\def\BibTeX{{\rm B\kern-.05em{\sc i\kern-.025em b}\kern-.08em
    T\kern-.1667em\lower.7ex\hbox{E}\kern-.125emX}}

\newcommand{\equref}[1]{(\ref{#1})}
\newcommand{\tabref}[1]{Table \ref{#1}}
\newcommand{\figref}[1]{Fig. \ref{#1}}

\def\th{\mathrm{th}}
\def\fix{\mathrm{fix}}

\begin{document}

\bstctlcite{BSTcontrol}
\title{Entropy Polarization-Based Data Compression Without Frozen Set Construction}
\author{Zichang~Ren, Yuping~Zhao}
\maketitle

\begin{abstract}
    Classical source polar codes require the construction of frozen sets for given sources. While this scheme offers excellent theoretical performance, it faces challenges in practical data compression systems, including sensitivity to the accuracy and computational complexity of the construction algorithm. In this letter, we explore the feasibility of construction-free polar compression schemes. By optimally selecting output symbols based on the decoder's behavior, the proposed scheme not only enhances flexibility but also achieves significant improvements in compression rates. Several enhancements are introduced to facilitate the practical implementation of the proposed scheme. Numerical results demonstrate the superior performance compared to existing polar compression approaches.
\end{abstract}

\begin{IEEEkeywords}
    Polar codes, source coding, data compression, successive cancellation
\end{IEEEkeywords}

\section{Introduction}

Since the introduction of information theory by Shannon \cite{6773024}, achieving the entropy of a source or the capacity of a channel has been a fundamental goal. Polar codes, proposed by Arikan \cite{5075875}, are the first to provably achieve channel capacity, and subsequent studies have enriched their theoretical framework, improved practical implementations, and led to their adoption in the 5G standard\cite{8962344}.

Beyond error correction, polar codes have also been explored for data compression. An initial attempt was made in \cite{5205860}, leveraging the duality between source coding and channel coding. Later, Arikan introduced entropy polarization theory in \cite{5513567}, proposing a fixed-length coding scheme based on frozen sets and the successive cancellation (SC) decoding algorithm, which provably achieves source entropy. Subsequent works, such as \cite{5437372}, \cite{6284254}, and \cite{5503220}, extended this approach to lossy compression and scenarios like Slepian-Wolf and Wyner-Ziv coding. Additionally, \cite{5513561} proposed an error-free polar compression scheme, which was later extended to non-binary sources and list decoding algorithms in \cite{6620403}. In recent years, there has been a growing body of work on joint source-channel polar coding, such as \cite{9453837} and \cite{10542342}.

However, existing polar compression schemes typically assume that frozen sets can be effectively constructed using established algorithms, such as those in \cite{5205857,6557004,8680016}. While these schemes demonstrate strong theoretical performance, they face significant challenges in practical data compression applications. For instance, modern image compression systems \cite{10378746,HUSSAIN201844} rely on existing entropy coding techniques\cite{arithmetic,duda_asymmetric_2009}, which are tailored to the varying statistical properties of different images. If polar codes were used for entropy coding in such scenarios, the need to construct separate frozen sets for each compression task would result in significant computational overhead and reduced flexibility. Furthermore, real-world source data often deviates from the assumption of identical distribution, which could lead to degradation in both the accuracy and computational efficiency of frozen sets derived from existing construction algorithms.

This letter aims to address these challenges and contribute to the practical applicability and flexibility of polar codes. Based on the observation on the behavior of the classical SC decoder with frozen sets, we introduce a data-dependent criterion for selecting the output symbols of the encoder. This strategy not only improves compression rates but also leads to a construction-free polar compression scheme. Furthermore, we present a series of practical enhancements that facilitate the immediate implementation of the proposed coding scheme. We believe our work offers a promising and insightful direction for the practical application of entropy polarization-based compression methods.

\section{Preliminaries}

\subsection{Notations}

We will focus exclusively on independently distributed discrete memoryless sources, denoted by $X$, and will discuss binary sources unless stated otherwise. A source realization $x \in \{0, 1\}$ represents the bit value output by the source in a specific experiment. For convenience, the subscript notation $X_{i:j}$ denotes the sequence $X_i, X_{i+1}, \dots, X_j$, with $i > j$ indicating an empty sequence. Similarly, when the subscript is a set, it refers to the sequence indexed by the elements of that set. For example, $x_{\mathcal{A}}$ represents the sequence $x_{i_1}, \ldots, x_{i_{\lvert \mathcal{A} \rvert}}$, where the indices $i_k \in \mathcal{A}$ are arranged in ascending order such that $i_k < i_{k+1}$. The notation $\lvert \mathcal{A} \rvert$ denotes the number of elements in the set $\mathcal{A}$. Throughout this letter, the symbol $\triangleq$ is used extensively to denote definitions.

\subsection{Source Polar Codes}

If we apply the polar transform to sources $X_{1:N}$ with $N = 2^n$ and $n \geq 1$:
\begin{equation}
    U_{1:N} = X_{1:N} \mathbf{G}_N,
    \label{equ:src_trans}
\end{equation}
where $\mathbf{G}_N = \mathbf{G}_2^{\otimes n}$ represents the $n$-th Kronecker power of $\mathbf{G}_2 = \begin{pmatrix} 1 & 0 \\ 1 & 1 \end{pmatrix}$, the conditional entropies:
\begin{equation}
    \mathcal{H}_i \triangleq H(U_i \mid U_{1:i-1}),
    \label{equ:define_H}
\end{equation}
will polarize. For any $\delta \in (0,1)$, as $N$ approaches infinity, the proportion of indices $i$ for which $\mathcal{H}_i \in (1 - \delta, 1]$ converges to the source entropy $H(X)$, while the proportion for which $\mathcal{H}_i \in [0, \delta)$ converges to $1 - H(X)$.

Based on this property, a lossless compression scheme is proposed in \cite{5513567}. For a given rate $R$, the frozen set $\mathcal{F}$ is defined to satisfy two conditions: $\lvert \mathcal{F} \rvert = \lceil NR \rceil$ and $\forall i \in \mathcal{F}, \forall j \notin \mathcal{F}, \mathcal{H}_i \geq \mathcal{H}_j$. The encoding process begins by applying the polar transform to the source realization $x_{1:N}$ to generate $u_{1:N} = x_{1:N} \mathbf{G}_N$, followed by the output of the subsequence $u_{\mathcal{F}}$. The decoder applies the SC decoding algorithm on $u_{\mathcal{F}}$ to estimate $\hat{u}_{1:N}$, and subsequently applies the inverse transform to recover the source realization as $\hat{x}_{1:N} = \hat{u}_{1:N} \mathbf{G}_N^{-1}$.

This is a fixed-length source coding technique, and for any $R>H(X)$, as $N$ tends to infinity, the decoding error rate approaches 0, thereby achieving the source entropy.

\section{Proposed Scheme}

\subsection{A Construction-Free Criterion}

The fundamental principle behind the freezing strategy is to provide the decoder with prior knowledge of a predetermined subset of sources, aiming to minimize the decoding error rate. Specifically, the task of a standard SC decoder is to compute the conditional distribution:
\begin{equation}
    p_i(\tilde{u} \mid u_{1:i-1}) \triangleq \Pr(U_i = \tilde{u} \mid U_{1:i-1} = u_{1:i-1}),
    \label{equ:cond_prob}
\end{equation}
and perform maximum-likelihood (ML) decision-making for the unknown symbols. Focusing on the entropy of this distribution, we denote it as $\hbar_i(u_{1:i-1})$, or simply $\hbar_i$, and observe the following equality:
\begin{equation}
    \mathcal{H}_i = \mathbb{E}(\hbar_i(u_{1:i-1})),
    \label{equ:h_vs_H}
\end{equation}
which implies that $\mathcal{H}_i$ represents the average entropy of $p_i(\tilde{u} \mid u_{1:i-1})$. For frozen indices $i \in \mathcal{F}$, where $\mathcal{H}_i$ is close to 1, the corresponding $p_i(\tilde{u} \mid u_{1:i-1})$ is, on average, close to a uniform distribution, leading to a higher ML error rate.

Our work is motivated by a slight mismatch between the freezing strategy and the SC decoding algorithm. Specifically, during the $i$-th step of SC decoding, the average information provided by symbol $u_i$ is actually $\hbar_i$, not $\mathcal{H}_i$. This mismatch, while commonly observed in polar coding research, is generally inconsequential to theoretical optimality, as indicated by equation \equref{equ:h_vs_H}. Furthermore, it remains unresolvable for channel polar codes, where the decoder receives codewords affected by unpredictable channel noise, and the encoder only knows the expected value $\mathcal{H}_i$, not the precise $\hbar_i$. However, this limitation does not apply to source polar codes. This subtle distinction allows the source encoder to replicate the decoder's behavior, directly obtaining $\hbar_i$, which offers two key advantages:

\begin{itemize}
    \item It potentially improves compression rates, as $\hbar_i$ not only reflects the statistical properties of $X_{1:N}$, but also takes into account the specific values of $x_{1:N}$ and the behavior of the decoder.
    \item It enables a construction-free scheme, as the encoder can obtain $p_i(\tilde{u} \mid u_{1:i-1})$ and $\hbar_i$ by performing a rate-1 standard SC decoding process on $u_{1:N}$, where $\mathcal{F} = \{1, \dots, N\}$, requiring no construction and avoiding the accuracy and complexity issues.
\end{itemize}

One challenge remains: $\hbar_i$ depends on the sequence $u_{1:N}$, which is unknown to the decoder prior to decoding. The following section will provide the corresponding solution.

\subsection{Basic Principles of the Proposed Scheme}\label{sec:basic}

We focus on error-free schemes. One of the primary tasks of the encoder is to convey the error set:
\begin{equation}
    \mathcal{E} \triangleq \left\{i : u_i \neq \mathop{\arg\max}\limits_{\tilde{u} \in \left\{0, 1\right\}} p_i(\tilde{u} \mid u_{1:i-1}) \right\},
    \label{equ:error_set}
\end{equation}
to the decoder. A straightforward approach would involve storing each index in $\mathcal{E}$ using $\log_2{N}$ bits, but this fails to achieve compression. Instead, we propose exploiting the entropy polarization property through a high-entropy set strategy. Specifically, we define a threshold $\hbar_{\th}$ as the minimum entropy corresponding to $\mathcal{E}$:
\begin{equation}
    \hbar_{\th} \triangleq \min\{\hbar_i : i \in \mathcal{E}\},
\end{equation}
and retain all indices with entropy no less than this threshold, forming the set:
\begin{equation}
    \mathcal{G} \triangleq \{i : \hbar_i \geq \hbar_{\th}\}.
    \label{equ:proposed_error_set}
\end{equation}

The encoder's output is $u_{\mathcal{G}}$ and $\hbar_{\th}$. Since $\mathcal{G}$ is unknown to the decoder, $u_{\mathcal{G}}$ can also be represented as $v_{1:L}$, where $L \triangleq \lvert \mathcal{G} \rvert$ denotes the code length.

With the assistance of the threshold $\hbar_{\th}$, the decoder can recover the set $\mathcal{G}$ using a sequential determination strategy. We introduce a pointer $\tau$ to track the symbol position in $v_{1:L}$, initialized to 1. The decoder proceeds sequentially from $i=1$ to $N$, computing $p_i(\tilde{u} \mid u_{1:i-1})$ and $\hbar_i$ at each step. Whenever $\hbar_i \geq \hbar_{\th}$, the current index $i$ is identified as part of $\mathcal{G}$. In this case, the value of $u_i$ is set to $v_{\tau}$, and the pointer $\tau$ is then incremented to the next position. If $\hbar_i < \hbar_{\th}$, the current index does not belong to $\mathcal{G}$, and the decoder makes an ML decision using $p_i(\tilde{u} \mid u_{1:i-1})$, leaving the pointer $\tau$ unchanged.

The proposed encoding and decoding algorithms are both with a computational complexity of $\mathcal{O}(N \log{N})$, consistent with classical schemes\cite{5513567}. They demonstrate the core principles of the proposed scheme and provide a foundational framework for further development. In the following sections, several enhancements are introduced to improve their practical applicability.

\subsection{Matching Principles}

\subsubsection{Non-standard SC Decoding Algorithm}

The widely used decoding algorithms for binary polar codes are those based on the min-sum algorithm and its variations\cite{8962344}. These algorithms improve hardware efficiency when computing the log-likelihood ratio (LLR):
\begin{equation}
    \gamma_i(u_{1:i-1}) \triangleq \log{\dfrac{p_i(0 \mid u_{1:i-1})}{p_i(1 \mid u_{1:i-1})}},
\end{equation}
but they sacrifice precision, which may lead to failure in correctly decoding $u_{\mathcal{G}}$. If we aim to leverage these non-standard algorithms to enhance computational efficiency in practical implementations, the encoding algorithm must be designed to match the decoder.

Specifically, when a non-standard SC decoding algorithm is employed by the decoder, the definition of the error set must be modified to align with the LLR-based decoding rule:
\begin{equation}
    \mathcal{E}_{\mathrm{LLR}} \triangleq \left\{i : u_i \neq \dfrac{1 - \mathrm{sign}(\gamma_i)}{2} \right\},
    \label{equ:error_set_mod}
\end{equation}
and the symbol selection criterion must be adjusted to retain symbols with smaller absolute LLR values (i.e., those with higher entropy):
\begin{equation}
    \mathcal{G}_{\mathrm{LLR}} \triangleq \{i : \lvert \gamma_i \rvert \leq \gamma_{\th}\},
    \label{equ:proposed_error_set_mod}
\end{equation}
where $\gamma_{\th} \triangleq \max\{\lvert \gamma_i \rvert : i \in \mathcal{E}_{\mathrm{LLR}}\}$.

These tasks should be executed by the encoder through a single rate-1 corresponding non-standard SC decoding process, in contrast to the standard SC algorithm.

\subsubsection{Non-binary Compression}

Another matching principle applies in non-binary scenarios. A more suitable symbol selection criterion, which better matches the decoder's behavior than non-binary entropy, is the ML error rate, defined as:
\begin{equation}
    \epsilon_i(u_{1:i-1}) \triangleq 1 - \max\limits_{\tilde{u} \in \left\{0, 1\right\}} p_i(\tilde{u} \mid u_{1:i-1}).
\end{equation}

It is also recommended to use $\epsilon$ instead of $\hbar$ in binary scenarios, because it enhances hardware efficiency by eliminating the need for logarithmic operations.

\subsection{Harnessing a Fixed Threshold}\label{sec:enhanced}

The basic form of the proposed scheme may incur additional storage overhead due to the variability of the threshold $\hbar_{\th}$ across different source realizations. This can be addressed by employing a fixed threshold, denoted as $\hbar_{\fix}$. The encoder outputs the sequence $u_{\mathcal{G}_{\fix}}$, where:
\begin{equation}
    \mathcal{G}_{\fix} \triangleq \left\{i : \hbar_i \geq \hbar_{\fix} \right\}.
\end{equation}

When $\hbar_{\th} \geq \hbar_{\fix}$ for a given source realization, $\mathcal{G}_{\fix}$ retains more symbols than necessary, which impacts the compression rate but does not lead to decoding failure. Conversely, if $\hbar_{\th} < \hbar_{\fix}$, some symbols in $\mathcal{G}$ are missing, resulting in decoding failure. Therefore, the encoder should additionally output:
\begin{equation}
    \mathcal{G}^*_{\fix} \triangleq \left( \mathcal{G} \setminus \mathcal{G}_{\fix} \right) \cap \mathcal{E},
    \label{equ:error_set_add}
\end{equation}
with each element being stored using $\log_2{N}$ bits. The decoder can correct these erroneous bits through bit-flipping.

From a practical perspective, both $u_{\mathcal{G}_{\fix}}$ and $\mathcal{G}^*_{\fix}$ are variable-length sequences. To distinguish between them, the decoder must know the length of at least one of these segments. Since $\lvert \mathcal{G}_{\fix} \rvert \leq N$, a fixed number of $\log_2{N}$ bits are sufficient for this purpose. In summary, a practical storage scheme would then output the tuple $(\lvert \mathcal{G}_{\fix} \rvert, u_{\mathcal{G}_{\fix}}, \mathcal{G}^*_{\fix})$.

For non-binary sources (with base-$r$ representation), in addition to $\mathcal{G}^*_{\fix}$, the decoder must also be provided with the corresponding correct values, as bit-flipping is no longer applicable. We propose outputting the modulo-$r$ difference between the ML-decided values and the true values, denoted as $\bar{u}_{\mathcal{G}^*_{\fix}}$. This introduces an additional storage overhead of $\log_r(r-1)$ symbols. The resulting storage scheme becomes $(\lvert \mathcal{G}_{\fix} \rvert, u_{\mathcal{G}_{\fix}}, \mathcal{G}^*_{\fix}, \bar{u}_{\mathcal{G}^*_{\fix}})$. Distinguishing between $\mathcal{G}^*_{\fix}$ and $\bar{u}_{\mathcal{G}^*_{\fix}}$ is straightforward, as the length ratio of the two segments is already known to be $\log_r{N}$ for $\mathcal{G}^*_{\fix}$ and $\log_r(r-1)$ for $\bar{u}_{\mathcal{G}^*_{\fix}}$.

Note that the above scheme also applies to the binary case, where $\bar{u}_{\mathcal{G}^*_{\fix}}$ is an all-1-bit sequence, requiring no storage and equivalent to bit-flipping. In summary, the compression rate for $r \geq 2$ when adopting a fixed threshold is given by:
\begin{equation}
    R_{\fix} = \dfrac{\log_r{N}}{N} + \lvert\mathcal{G}_{\fix}\rvert \dfrac{1}{N} + \lvert \mathcal{G}^*_{\fix}\rvert \dfrac{\log_r{N}+\log_r(r-1)}{N}.
    \label{equ:rate_fix}
\end{equation}

Inspired by previous works \cite{5513561} and \cite{6620403}, we set:
\begin{equation}
    \epsilon_{\fix} = \frac{1}{\log_r{N} + \log_r(r-1)},
    \label{equ:threshold_fix}
\end{equation}
to achieve the optimal compression rate. For other symbol selection criteria, conversions can be made based on their relationships, such as $\gamma_{\fix} = \lvert \log(1 - \epsilon_{\fix}) / \log \epsilon_{\fix} \rvert$.

At this stage, all practical enhancement strategies have been presented. We adopt this fixed ML error rate threshold-based strategy as the final version of the proposed scheme.

\begin{table*}[t]
    \centering
    \caption{Average code rates for different binary sources, compared with error-free scheme presented by \cite{5513561}.}
    \label{tab:test_02_01}
    \renewcommand{\arraystretch}{0.5}
    \begin{tabularx}{\textwidth}{*{13}{c}}
        \toprule
        \textbf{Sources} & \textbf{Schemes} & $N=2^{8}$ & $N=2^{9}$ & $N=2^{10}$ & $N=2^{11}$ & $N=2^{12}$ & $N=2^{13}$ & $N=2^{14}$ & $N=2^{15}$ & $N=2^{16}$ & $N=2^{17}$ \\
        \midrule
        \multirow{2}{*}{$\mathbf{0.1}$} & \textbf{Proposed} & 0.13846 & 0.12541 & 0.11937 & 0.11345 & 0.11088 & 0.10891 & 0.10794 & 0.10672 & 0.10592 & 0.10554 \\
        \cmidrule(lr){2-2} \cmidrule(lr){3-12}
        & \textbf{Error-free} & 0.16305 & 0.15451 & 0.15122 & 0.14293 & 0.13613 & 0.13014 & 0.12586 & 0.12150 & 0.11813 & 0.11565 \\
        \midrule
        \multirow{2}{*}{$\mathbf{0.5}$} & \textbf{Proposed} & 0.56154 & 0.54844 & 0.53636 & 0.52948 & 0.52556 & 0.52114 & 0.51845 & 0.51606 & 0.51309 & 0.51144 \\
        \cmidrule(lr){2-2} \cmidrule(lr){3-12}
        & \textbf{Error-free} & 0.59820 & 0.58459 & 0.57271 & 0.56381 & 0.55428 & 0.54897 & 0.54266 & 0.53716 & 0.53184 & 0.52716 \\
        \midrule
        \multirow{2}{*}{$\mathbf{0.9}$} & \textbf{Proposed} & 0.94939 & 0.93465 & 0.92417 & 0.91769 & 0.91437 & 0.91072 & 0.90944 & 0.90755 & 0.90640 & 0.90562 \\
        \cmidrule(lr){2-2} \cmidrule(lr){3-12}
        & \textbf{Error-free} & 0.94133 & 0.93833 & 0.93306 & 0.92839 & 0.92604 & 0.92109 & 0.91952 & 0.91653 & 0.91446 & 0.91252 \\
        \bottomrule
    \end{tabularx}
\end{table*}

\begin{table*}[t]
    \centering
    \caption{Average code rates for different ternary sources, compared with error-free scheme presented by \cite{6620403}.}
    \label{tab:test_02_02}
    \renewcommand{\arraystretch}{0.5}
    \begin{tabularx}{\textwidth}{*{13}{c}}
        \toprule
        \textbf{Sources} & \textbf{Schemes} & $N=2^{8}$ & $N=2^{9}$ & $N=2^{10}$ & $N=2^{11}$ & $N=2^{12}$ & $N=2^{13}$ & $N=2^{14}$ & $N=2^{15}$ & $N=2^{16}$ & $N=2^{17}$ \\
        \midrule
        \multirow{2}{*}{$\mathbf{0.3}$} & \textbf{Proposed} & 0.35516 & 0.34399 & 0.33684 & 0.32990 & 0.32520 & 0.32192 & 0.31837 & 0.31558 & 0.31310 & 0.31151 \\
        \cmidrule(lr){2-2} \cmidrule(lr){3-12}
        & \textbf{Error-free} & 0.38578 & 0.37314 & 0.36580 & 0.35572 & 0.34873 & 0.34224 & 0.33620 & 0.33063 & 0.32605 & 0.32257 \\
        \midrule
        \multirow{2}{*}{$\mathbf{0.5}$} & \textbf{Proposed} & 0.56134 & 0.54867 & 0.54055 & 0.53295 & 0.52783 & 0.52297 & 0.52004 & 0.51703 & 0.51432 & 0.51237 \\
        \cmidrule(lr){2-2} \cmidrule(lr){3-12}
        & \textbf{Error-free} & 0.59158 & 0.58047 & 0.56924 & 0.55991 & 0.55310 & 0.54469 & 0.53872 & 0.53344 & 0.52816 & 0.52428 \\
        \midrule
        \multirow{2}{*}{$\mathbf{0.8}$} & \textbf{Proposed} & 0.85015 & 0.83727 & 0.83088 & 0.82430 & 0.81981 & 0.81648 & 0.81409 & 0.81181 & 0.81013 & 0.80845 \\
        \cmidrule(lr){2-2} \cmidrule(lr){3-12}
        & \textbf{Error-free} & 0.85863 & 0.85146 & 0.84705 & 0.83967 & 0.83476 & 0.82981 & 0.82598 & 0.82234 & 0.81915 & 0.81619 \\
        \bottomrule
    \end{tabularx}
\end{table*}

\section{Numerical Results}

In this section, we validate the feasibility of the proposed scheme, with each data point simulated across 1000 iterations unless stated otherwise.

\begin{figure}[t!]
    \centerline{\includegraphics[width=0.45\textwidth]{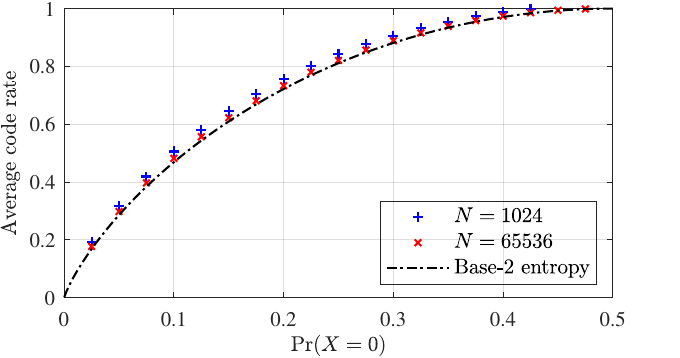}}
    \caption{Average code rates for various binary sources $X$.}
    \label{fig:test_01_03}
\end{figure}

Firstly, we simulate the average compression rates of the proposed scheme for several binary sources with entropies ranging from 0 to 1, corresponding to bit-0 generation probabilities ranging from 0 to 0.5, at code lengths of $N=1024$ and $N=65536$. As shown in \figref{fig:test_01_03}, the results demonstrate that, at moderate code lengths, the compression rate is already very close to the source entropy.

We then compare the average compression rates of the proposed scheme with the error-free polar compression scheme presented in \cite{5513561}. Using a similar experimental setup, simulations were conducted for binary sources with base-2 entropies of 0.1, 0.5, and 0.9, under various code lengths. The results shown in \tabref{tab:test_02_01} demonstrate that the proposed scheme achieves significantly better compression rates, except in the case with an entropy of 0.9 and a code length of 256. This anomaly is likely due to the storage overhead of $\lvert \mathcal{G}_{\fix} \rvert$ at very short code lengths.

As for non-binary scenarios, we compare the proposed scheme with the method presented in \cite{6620403}. Simulations were conducted for ternary sources with base-3 entropies of 0.3, 0.5, and 0.8, corresponding to the distributions $\{0.9214, 0.0393, 0.0393\}$, $\{0.07, 0.09, 0.84\}$, and $\{0.1, 0.275, 0.625\}$, respectively, under different code lengths. The results are presented in \tabref{tab:test_02_02}.

The superior performance of the proposed scheme compared to existing methods can be attributed to the fact that it accounts for both the specific values of source realizations and the specific behavior of the SC decoding process. This approach is more precise than the classical freezing strategy, which relies solely on conditional entropies that depend only on the statistical properties of the source.

\begin{figure}[t!]
    \centerline{\includegraphics[width=0.45\textwidth]{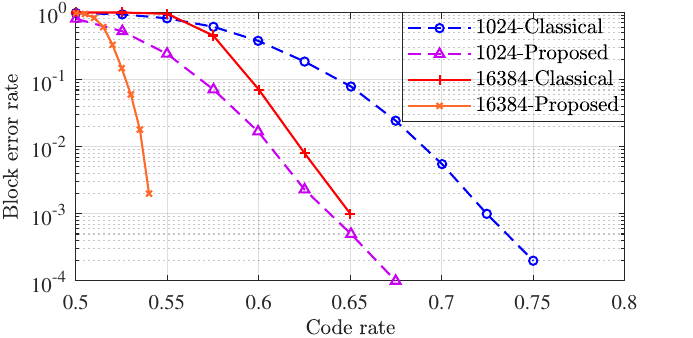}}
    \caption{The rate-distortion performance compared with classical scheme\cite{5513567}.}
    \label{fig:test_03_01}
\end{figure}

Additionally, one may be interested in comparing the performance of the proposed scheme with that of the classical source polar codes \cite{5513567}, but a key distinction is that the latter are fixed-length compression schemes, with performance typically measured by the block error rate (BLER) at a given code rate, whereas the proposed scheme is error-free, making BLER an inapplicable metric. To enable a fair comparison without altering the classical scheme, we propose a fixed-length scheme by truncating the output of the proposed scheme to a length no greater than $NR$, thus transforming it into a fixed-length scheme with rate $R$. For source realizations with $R_{\text{fix}} > R$, decoding will fail; otherwise, decoding is guaranteed to succeed. Simulations were conducted for a binary source with entropy 0.5, under code lengths of $N=1024$ and $N=16384$, across 10,000 simulation iterations. The results, as shown in \figref{fig:test_03_01}, demonstrate that when BLER is used as the distortion measure, the proposed scheme significantly outperforms source polar codes in terms of rate-distortion performance.

\section{Conclusion}

This letter presents a novel construction-free polar coding scheme, which we believe offers a promising and insightful direction for applying entropy polarization-based coding to real-world data compression systems. The proposed encoder optimally selects output symbols based on decoder behavior, while the decoder utilizes the sequential nature of the SC decoding algorithm to determine the indices of each bit in the encoded sequence. To further enhance the scheme's applicability, we introduce matching principles for adapting to non-standard decoding algorithms and non-binary scenarios, along with a fixed threshold to facilitate immediate implementation. Simulation results demonstrate the proposed scheme's superior compression performance compared to existing polar compression methods.

\clearpage

\bibliographystyle{IEEEtran}
\bibliography{refs}

\end{document}